\documentclass[twocolumn,aps,pre,amsmath,showpacs]{revtex4}
\usepackage{graphicx}
\usepackage{dcolumn}%
\usepackage{bm}%
\usepackage{amssymb}
\usepackage{amsmath}
\usepackage{color}
\newcommand{\kb}{k_{_{\rm B}}}

\def\be{\begin{equation}}
\def\ee{\end{equation}}

\begin{document}

\title{Kramers rate theory of bound-state dissociation at low and high densities}
\author{A. Zaccone and E. M. Terentjev}
\affiliation{Cavendish Laboratory, University of Cambridge, JJ Thomson Avenue,
Cambridge CB3 0HE, U.K.}
\date{\today}
\begin{abstract}
\noindent
Calculating the microscopic dissociation rate of a bound state,
such as a classical diatomic molecule, has been difficult so far. The
problem was that standard theories require an energy barrier over which the
bound particle (or state) escapes into the preferred low-energy state. This is
not the case when the
long-range repulsion responsible for the barrier is either absent or screened
(as in Cooper pairs, ionized plasma, or biomolecular complexes). We solve this
classical problem by accounting
for entropic memory at the microscopic level. The theory predicts
dissociation rates for arbitrary potentials and is successfully tested
on the example of plasma, where it yields an estimate of ionization in the
core of Sun in excellent agreement with experiments. In biology, the new theory
accounts for crowding in
receptor-ligand kinetics and protein aggregation.
\end{abstract}

\maketitle

\begin{figure}
\includegraphics[width=0.9\linewidth]{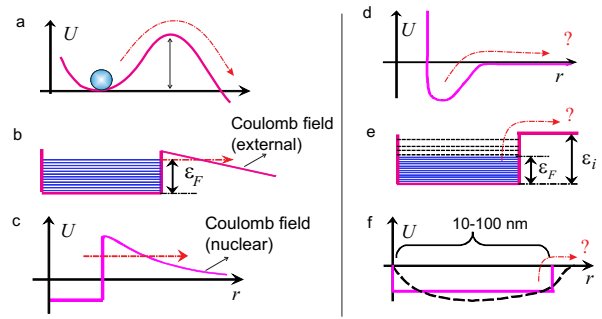}
\caption{Examples of situations where the dissociation problem
is well defined (a-c) and where it is not (d-f). (a) The Kramers
escape problem~\cite{kramers}. (b) Field-emission of electrons~\cite{Born},
where the external electric field gives rise to a well-defined
barrier through which the electron
can escape or tunnel away. (c) The escape of $\alpha$-particles from atomic
nuclei~\cite{Born}, where the repulsion is between positive charges. (d) The
internuclear
potential of a diatomic molecule or molecular complex~\cite{Israelachvili}. (e)
Potential of electrons
at metal surfaces (thermionic effect) or by neutrons evaporating from hot
nuclei~\cite{Fermi}. (f) The attractive potential between the two
electrons of a Cooper pair in real-space~\cite{Weisskopf}.}
\label{fig1}
\end{figure}

The rate of escape of a classical particle over an energy barrier is a
well-posed problem as long as the potential energy features a barrier or
transition-state that has to be crossed~\cite{Haenggi}. This is the classical
Kramers problem~\cite{kramers},
Fig.~\ref{fig1}(a), that has served very well in many areas of science.
However,
the escape-rate problem is known to be ill-defined when the particle is
trapped in a potential well which is the only point of minimum in the potential
profile (which then either diverges or reaches asymptotes along
the coordinated axis)~\cite{Haenggi}. The latter case is typically
exemplified by two noble gas atoms bonded by van der Waals forces such that
each atom is trapped in the Lennard-Jones potential well.
The same applies to many other sorts of bound states (e.g. diatomic
molecules~\cite{Born,Schollkopf}, deuterons~\cite{Fermi}, Cooper
pairs~\cite{Weisskopf}, nuclear neutrons~\cite{Fermi}, etc.) in the absence of
long-range repulsion,
Fig.~\ref{fig1}(d)-(f). In all such cases, the Kramers and transition-state
theories
cannot be applied because the flux driving the particle out of the
well cannot be defined, and in fact not even an energy barrier to be
crossed can be identified. This is in contrast to many other situations
where the competing long-range repulsion
gives rise to a well-defined barrier, Fig.~\ref{fig1}(a)-(c).
In spite of this theoretical difference,
in all these systems the observed time of escape is still finite and a
theoretical estimate is desirable for many applications. This is a
long-standing and well-known problem~\cite{Haenggi}. The common remedy to this
difficulty, so far, was to employ fictitious absorbing boundaries or long-range
repulsions to artificially create a barrier or transition-state which would
allow one to apply the
Kramers theory for the escape rate over such a
barrier~\cite{McLeish}. This procedure has the shortcoming that the
resulting barrier or transition-state is completely arbitrary, and so
are the results for the calculated dissociation rates.

Inspired by Peierls~\cite{peierls}, here we propose a solution to this problem
by considering the role of
entropy in the escape process. {It is known that entropic contributions can
influence kinetics in several contexts~\cite{Feil}.
As an example we recall the problem of the equilibrium flux through bottlenecks
with a cross-section fluctuating in time studied by Zwanzig~\cite{Zwanzig},
where even in the absence of energy barriers due to conservative potentials,
the passage time is controlled by entropic barriers due to the fluctuations.
However, the question about the fundamental mechanism by which entropy comes
into play has remained largely unanswered and an analytical framework of
general applicability is lacking.
Here we show how entropy drives the flux responsible for dissociating the bound
states in the absence of stabilizing barriers} and analytically calculate the
microscopic dissociation rate of classical bound states. The generic theory is
validated by showing that it is able to recover the classical Saha ionization
degree in the dilute limit. Further, our approach is fully microscopic, which
allows us time to include the density effects in the recombination
process.

{\it Effective pair-interaction energy for dissociation.}  \
We start with two thermal particles bonded by an attractive interaction where
it is evident that upon moving apart from one another along the
radial coordinate
the two particles will be less favored in terms of potential energy but they
will gain a larger free volume, and hence more entropy. Furthermore, one should
remark that the ``pair-interaction energy $w(r)$'' between two molecules or
particles separated by a distance $r$ is usually identified by the force $f$
acting between the two particles via $f=-\textrm{d}w(r)/\textrm{d}r$; hence one
finds the work that can be done by the force. This means that $w(r)$ is
actually the \emph{free energy} or the available energy for the two-particle
system~\cite{Israelachvili} because the local collisional physics allows us to define an entropy in addition to the potential energy.
Rigorously, the effective pair-potential for the two-particle system is thus
given by: $w(r)=U(r)-T\Delta S(r)$, where $U$ is the conservative potential
energy of interaction between
the two particles and $\Delta S$ is the mixing entropy of the system. Let us
interpret the dissociation coordinate $r$ as a time-averaged position where the
average is taken over a time $t>\tau_{c}$ where $\tau_{c}$ is the collision
time-scale between the two bound particles. Then, the contact force $f_{c}$ due
to collisions between the two particles {constrained to remain at close contact
over a finite amount of time} obeys the scaling relation~\cite{Wyart}:
$f_{c}\sim k_{B}T/r$.
For two hard-spheres constrained into a small portion of space where they
collide repeatedly with each other, the contact force can be integrated to give
an associated pair-interaction free energy which is related to the entropy of
the two-particle system: $T\Delta S(r)\sim k_{B}T\ln (r/R)$ where we chose the
integration constant equal to $-\ln R$. In the presence of a conservative
interaction $U(r)$ the total interaction force is thus given by
$f=f_{c}-(dU(r)/dr)$ and the corresponding total effective pair-interaction is
given by
$w(r)\simeq U(r)-k_{B}T\ln (r/R)$. A more precise form for this effective pair
interaction, including prefactors, can be derived using a different method,
which makes use of the Onsager excluded volume theory~\cite{Kleman}. With this
method one obtains:
\begin{equation}
\Delta
S=k_{B}\left(2\ln\frac{V_{1}}{2v_{p}}+\frac{u}{2v_{p}}\right)=
k_{B}\left(2\ln\frac{r^{3}}{2R^{3}}+4\right),
\label{eqmod5}
\end{equation}
where $V_{1}(r)=4\pi r^{3}/3$, $u=4\pi(2 R)^{3}/3$, and $v_{p}$ is the volume
of one
particle. This equation expresses that fact that upon moving the particles
apart along
the outward radial coordinate $r$ there is a net entropy gain which arises from
the increased number of degrees of freedom explored by the particles. Hence the
interaction energy can be rewritten as
\begin{equation}
\begin{aligned}
w(r)=U(r)-k_{B}T\left(6\ln\frac{r}{R}+4-2\ln2\right).
\label{eqmod6}
\end{aligned}
\end{equation}

Based on these considerations, the effective pair-interaction potential given
by Eq.~(\ref{eqmod6}) is coarse-grained in the sense that is valid only for
particles that have been kept at close distance for a time $t\gg
\tau_{c}$~\cite{Wyart}. As such, Eq.~(\ref{eqmod6}) can be applied to the
dissociation of pairs of particles which are bonded by some attractive
potential $U(r)$. On the other hand, for two particles which approach each
other from far apart, their mutual interaction prior to colliding happens on a
time scale $t\leq\tau_{c}$, i.e. they cannot explore their mutual excluded
volume.
Therefore for the
recombination process there is no entropic effect and $w(r)\rightarrow U(r)$.
This can be understood by recalling that entropy in classical systems is
ultimately related to collisions which are important in the bound-state where the particles preserve ``memory" of each
other~\cite{Frey} over the life-time of the bound state and this makes their configurational
entropy depend on their separation. In the recombination process, however, since the conservative potentials $U$ that we consider here do not have a barrier, the recombination of diffusive particles is diffusion-limited and occurs at the very first collision event between the two particles. Therefore, no entropic contribution applies to recombination processes of this kind where the recombination rate is uniquely determined by the diffusion process in the field of the conservative potential $U$.
%
%
In the following we will use these arguments to derive microscopic dissociation
and recombination rates with the tools of statistical mechanics.

\begin{figure}
\includegraphics[width=0.6\linewidth]{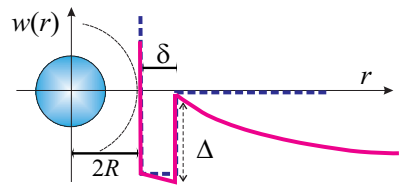}
\caption{Schematic of the rectangular-well potential around a particle of
radius $R$
used for $U(r)$ in the derivation, shown by the dashed line.
It is the effective potential (free energy)
$w(r)$, solid line, with its entropic correction making the higher separation
$r$ favorable, that affects the dissociation of the bound state.}
\label{fig3}
\end{figure}

{\it Dissociation rate.} \
Without loss of generality, but to simplify the algebra and
emphasize the qualitative point, let $U(r)$ in Eq.~(\ref{eqmod6})
be a rectangular-well attractive
potential, of range $\delta$ (Fig.~\ref{fig3}, dashed line), so that two
particles constrained within a distance $r=2R+\delta$.
Due to the entropy contribution, $w(r)$ from Eq.~(\ref{eqmod6})
 decreases logarithmically at large
distances, while having little effect on the proximity well.
Such an effective potential (free energy)
features a now well-defined barrier over which the bound particle can escape.
Using the Kramers escape theory for the 3D spherically-symmetric
problem~\cite{Haenggi}, the escape time, i.e. the reciprocal of the escape rate
$\kappa$, is given by
\begin{equation}
\frac{1}{\kappa_{diss}}
=\int_{2R}^{r_{\mathrm{min}}}\frac{e^{-w(r)/k_{B}T}}{D}r^{2}dr
\int_{r_{\mathrm{\mathrm{min}}}}^{C}\frac{e^{w(r)/k_{B}T}}{r^{2}}dr .
\label{eqmod7}
\end{equation}
The upper limit $C\gg R+\delta$ is some arbitrary point far away along the
radial axis and
$r_{\mathrm{min}}=2R+\delta$ is the minimum of the effective potential $w(r)$;
$D$ is the diffusion
coefficient for this problem. Using Eq.~(\ref{eqmod6}) in Eq.~(\ref{eqmod7})
the integrals can be
evaluated analytically which gives the compact expression for the dissociation
rate:
\begin{equation}
\kappa_{diss}=  63
D\left[\frac{(2R+\delta)^{7}}{(2R+\delta)^{9}-(2R)^{9}}\right]e^{-\Delta/k_{B}T}.
\label{eqmod8}
\end{equation}
The escape rate is directly proportional to the diffusion coefficient, as in
the classical Kramers theory. It is important that this dissociation rate
decreases with the increasing width of the attractive well, $\delta$:
this is a natural reflection of the
fact that the effective frequency of the particle in the potential well
(related to the `rate of attempts' to overcome the barrier) increases as
$\delta \rightarrow 0$.
The height of the effective energy
barrier in the exponential thermal-activation term, is given by
$\Delta/k_{B}T$, that is, by the depth of the original square well potential.
This explains
why a number of {\it ad hoc} theories produce results that are qualitatively
valid; the crux here is in the detail.

{\it Recombination rate.} \
Once the dissociation of the bound state occurs, i.e.
the bound particle has crossed the barrier in the effective potential of
Fig.~\ref{fig3}, the two particles move apart. {For a dilute thermal system,
the
mechanism of recombination back into the bound state is controlled by the
diffusive transport. With interacting particles, the recombination rate
is determined by solving the stationary Smoluchowski (diffusion) equation in
the field of force due to the interaction. For the attractive square-well case
the solution is:}
\begin{equation}
{\kappa _{rec}} = \frac{{4\pi D}}{{{e^{ - \Delta /k_{B}T}}
\left( {\frac{1}{{2R}} - \frac{1}{{2R + \delta }}} \right) + \frac{{1}}{{2R +
\delta }}}}
\label{eqmod9}
\end{equation}
{Since $\delta$ is finite and for $\Delta/k_{B}T>1$, the first term in the
denominator is small compared
to the second term due to the exponential factor, and thus we obtain: }
\begin{equation}
\kappa_{rec} \approx 4\pi D (2R+\delta).
\label{eqmod10}
\end{equation}
{This result can be extended to denser fluids using a generalized
diffusion coefficient given by $D(n)=(D /\kb T) d\Pi (n)/dn $ where $\Pi$
is the osmotic pressure of the fluid medium and $n\equiv N/V$ is the average
number density of diffusing particles~\cite{Foffi}.}
Using this $D(n)$ in
the above solution to the Smoluchowski diffusion equation, one obtains the
generalization of
the recombination rate for dense and crowded systems~\cite{Foffi}:
$\kappa_{rec} \approx 4\pi D (2R+\delta)\Pi (n)/n \kb T$.
The correction can be easily evaluated once the equation of state (EOS)
of the fluid is known. In the simplest non-ideal case one can express
the EOS in terms of the virial expansion~\cite{McCammon}.
For model hard-sphere fluids one can use the Carnahan-Starling
EOS~\cite{Hansen_book}: $\Pi/n k_{B}T \equiv
Z(\phi)=(1+\phi+\phi^{2}-\phi^{3})/(1-\phi)^{3}$, where $\phi$ is the fraction
of occupied volume in the system, e.g. $\phi=(4\pi/3) R^{3}n$.
This leads to the recombination-rate coefficient
$ \kappa_{rec} = 4\pi D (2R+\delta)Z(\phi)$, which would be relevant for a
crowded environment. For a dilute solution of biomolecules which bind to form
complexes (e.g.
receptor-ligand binding) in a sea of other particles of similar size with which
they interact
by steric repulsion, this leads to recombination rates several times larger
than the Smoluchowski estimate.
{We remark that for $\phi\gtrsim0.40$, it is no longer safe to neglect entropic
effects on
recombination, due to the short-range liquid-like structuring (inhomogeneous
local density)
around the reactants which cause the incoming particle
to collide with an increasing amount of crowders before colliding with the
second reactant. This clearly leads to an entropic repulsion upon
recombination, as discussed in~\cite{Foffi}. These considerations fix the upper
validity limit of our theory to $\phi\simeq0.45$. Extension to glassy systems
is non-trivial and will be the object of future work.}

If the dense fluid is made of charged particles, such as in high-density
plasmas,
then the transport is
no longer diffusive but drift-controlled. In this case the oppositely charged
particles attract each other via the Coulomb force $\sim (Ze)^{2}/r^{2}$, where
$Ze$ is the electric charge of the particles. The crossover from
diffusion-controlled to drift-controlled transport typically happens when
$n[(Ze)^{2}/k_{B}T]\sigma \gg1$, where $\sigma$ is the collision cross-section.
In this case the total current of oppositely-charged particles entering the
interaction volume is given by: $4\pi r^{2} v_{d} n$, where
$v_{d}=(Ze/r^{2})(\mu_{-}+\mu_{+})$ is the drift velocity in the Coulomb
attraction field of approaching particles and $\mu$ the
electrical mobility. The electron-ion recombination-coefficient for dense
plasmas is thus given by $\kappa_{rec} \approx  4\pi  e\mu_{-}$,
assuming that the electron mobility $\mu_{-}$ is much greater than the ion
mobility and $Z=1$. This is a well-known result from the kinetic theory of
plasmas~\cite{Smirnov}.

{\it Dissociation equilibrium.} \
The equilibrium of association and dissociation reactions is of extreme
importance for \emph{in vivo} physiological processes, e.g. enzymatic activity
based on receptor-ligand complexes or protein
aggregation~\cite{Minton,Knowles}. We will give a treatment
of the general case of two classical particles bound by an attractive potential
which can
be directly applied to biomolecules. The net
attractive potential $U(r)$ is of the kind of Fig.~\ref{fig1}(d).
Given these conditions, the
chemical equilibrium constant $K$ for the dissociation-recombination process
$B+C\rightleftharpoons BC$ is given by the law of mass action as
\begin{equation}
K
=\frac{\kappa_{diss}}{\kappa_{rec}}=\frac{n_{B}n_{C}}{n_{BC}}=\frac{N_{B}N_{C}}{N_{BC}}\frac{1}{V}.
\label{eqmod13}
\end{equation}
If $N_{0}=N_{BC}+N_{B}=N_{BC}+N_{C}$ is the total number of each species
present in the system, the total number of particles, both bound and
dissociated, is given by $N=N_{0}(1+\alpha)$, where
$\alpha=N_{B}/N_{0}=N_{C}/N_{0}$ is the ``degree of
dissociation''~\cite{Landau}.
The relations for the molar fractions of the
various species follow at once: $N_{B}/N_{0}=\alpha/(1+\alpha)$
 and $N_{BC}/N_{0}=(1-\alpha)/(1+\alpha)$.

Using Eq.~(\ref{eqmod8}) and the high-density extended Smoluchowski rate
for $\kappa_{diss}$ and $\kappa_{rec}$, respectively, this leads to the
following expression for the dissociation degree in equilibrium:
\begin{equation}
\frac{\alpha^{2}}{1-\alpha^{2}}=\frac{63}{4\pi}\frac{(2R+\delta)^{6}}{(2R+\delta)^{9}
-(2R)^{9}}\frac{e^{-\Delta/k_{B}T}}{nZ(\phi)}.
\label{eqmod14}
\end{equation}
In the limit of long-range attraction, $\delta \gg 2R$, this simplifies
further and we get the following expression for the degree of dissociation
$\alpha$:
\begin{equation}
\alpha \approx
\left[1+(4\pi/63)nZ(\phi)\delta^{3}e^{\Delta/k_{B}T}\right]^{-1/2}.
\label{eqmod15}
\end{equation}
In biological systems one often encounters situations where the interaction
range can
be significant, such as in the ubiquitous case of hydrophobic
attraction~\cite{Israelachvili_Nature}. Equation~(\ref{eqmod15}) shows that the
dissociation degree in cases of biological relevance can have a sensitive
dependence upon the interaction range $\delta$, which has been neglected
in previous theories. Furthermore, Eq.~(\ref{eqmod15}) takes into account the
enhanced recombination in crowded systems.
In the opposite limit of short-range attraction (sticky or
adhesive particles~\cite{Frenkel}), $\delta\ll 2R$, one obtains:
\begin{equation}
\alpha \approx \left[1+ 7.2 nZ(\phi)R^{2}\delta e^{\Delta/k_{B}T}\right]^{-1/2}
\label{eqmod16}
\end{equation}
with a much weaker dependence on $\delta$.
{Finally, we should remark that in the above treatment we assumed
$n_{B}=n_{C}$,
which describes dissociation and ionization kinetics.
In receptor-ligand kinetics, however, the concentration of ligands normally
overwhelms the concentration of receptors, i.e. $n_{C}\gg n_{B}$.
Since we derived \emph{intrinsic} rates, we can apply our theory to any kind of
kinetics,
including the pseudo-first-order situation $n_{C}\gg n_{B}$. In this limit, the
kinetics is of the Langmuir type~\cite{Gaster}: $\theta  = K{n_C}/(1 +
K{n_C})$,
where $\theta=N_{BC}/(N_{B}+N_{BC})$, and $K$ is still given by
Eq.~(\ref{eqmod13})
with the rates for dissociation and association derived here.}

{\it Ionization equilibrium in plasmas.} \
We can further test this theory by addressing the problem of the thermal
dissociation of atoms (ionization). This problem was
famously addressed in the 1920's by Saha~\cite{Landau} by combining the
statistical
mechanics of the ideal gas with the chemical equilibrium (detailed balance)
assumption. By considering the chemical equilibrium of the ionization reaction
$A \rightleftharpoons I^{+} +e^{-}$, the chemical equilibrium
constant of the reaction is given by the same expression as
Eq.~(\ref{eqmod13}).
In the dilute limit, Eq.~(\ref{eqmod10}) for $\kappa_{rec}$ applies,
and using Eq.~(\ref{eqmod8}) for $\kappa_{diss}$ we obtain
\begin{equation}
\frac{\alpha^{2}}{1-\alpha^{2}}=\frac{63}{4\pi}\frac{V}{N v}e^{-\Delta/k_{B}T}
\label{eqmod17}
\end{equation}
where $N=N_{A}+N_{I^{+}}+N_{e^{-}}$. The interaction volume, as we have shown
above, is given by
$v\equiv[(2R+\delta)^{9}-(2R)^{9}]/(2R+\delta)^{6}$. The extension of
the radial probability amplitude
for a bound electron, $\lambda_{e^{-}}$, is approximately given by the width of
the attractive well, and therefore we take $\delta
\simeq \lambda_{e^{-}}$. In the classical approximation valid at high
temperatures, $\lambda_{e^{-}}$ is the thermal de Broglie wavelength,
$\lambda_{e^{-}}=\hbar/\sqrt{m_{e} k_{B}T/2\pi}$.
Furthermore, one has $\lambda_{e^{-}} \gg \lambda_{I^{+}}$, such that
$v\equiv[(\lambda_{I^{+}}+\lambda_{e^{-}})^{9}-(\lambda_{I^{+}})^{9}]/
(\lambda_{I^{+}}+\lambda_{e^{-}})^{6}\simeq\lambda_{e^{-}}^{3}$,
i.e. the same limit as used in Eq.~(\ref{eqmod15}).
With this replacement in Eq.~(\ref{eqmod17}),
our theory for a dilute plasma where free diffusion is the main transport
mechanism correctly reduces to the Saha equation~\cite{Landau}, apart from a
numerical
pre-factor of order unity. This result demonstrates the validity of the proposed calculation
scheme.

The Saha equation works well for dilute plasmas, but it breaks
down in the core of the stars~\cite{Salaris}. In fact, the opacity in the interior of stars (such as the Sun) is extremely low for that high density, and this implies that the interior is formed by completely ionized matter with
$\alpha\simeq1$. Evaluating the Saha equation at the typical conditions of the
stellar interiors gives $\alpha\simeq0.7-0.8$ that is incompatible with the
observed vanishing opacity.
We can now extend the theory to the high-density regime because we can
include transport effects, that are important at high density, via the
recombination rate which enters in our model. Using the recombination rate for
high-density plasmas given by  $\kappa_{rec} \approx  4\pi  e\mu_{-}$ in the
equilibrium
constant for atomic dissociation $K = \kappa_{diss}/\kappa_{rec}$, we obtain
\begin{equation}
\frac{\alpha^{2}}{1-\alpha^{2}}=\frac{63}{4\pi}\frac{k_{B}T \, V}{e^{2}
\lambda_{e^{-}}^{2} N}e^{-\Delta/k_{B}T}
\label{eqmod18}
\end{equation}
where we used the Einstein relation $eD/k_{B}T=\mu$ for the
single-electron ionization, $Z=1$, and continued using $\delta \simeq
\lambda_{e^{-}}$ for the bound state of electron.
Let us calculate $\alpha$ for the conditions corresponding to the core of the
Sun, i.e. $T=10^{7}K$, $N/V=10^{26}\textrm{cm}^{-3}$, and assume pure hydrogen.
The Saha equation gives $\alpha\simeq 0.79$, whereas from Eq.~(\ref{eqmod18})
we obtain a much more realistic
$\alpha\simeq0.99$. Clearly, Eq.~(\ref{eqmod18}) is in excellent
agreement with the experimental data indicating fully-ionized
plasma conditions in the core of the Sun~\cite{Mullan}.

To conclude, we have proposed a theoretical
scheme that allows one to calculate the intrinsic dissociation rate of bound
states with purely attractive potentials, something so far
possible only for potentials that feature long-range repulsion competing with
attraction. In all the other cases, including dissociation of neutral atoms,
molecules, Cooper pairs, molecular complexes, this was not possible
and required various {\it ad hoc} assumptions.
We validated our theory for its predictions of the ionization degree of
hydrogen-like atoms, a problem relevant in several branches of astrophysics.
Our
theory correctly reproduces the Saha formula at steady-state
in the ideal-gas limit. Further, it can be easily extended to deal with
the high-density limits. In fact it
predicts full ionization in the core of stars, thus dramatically improving
over the Saha equation that predicts partial ionization.
Further applications are in biology where it allows one to
calculate dissociation equilibria of receptor-ligand or protein complexes in
presence of cellular crowding.

{\it {\bf Acknowledgments.}} \ We are grateful for discussions and input of
M. Warner and A. Lasenby. This work has been supported by the EPSRC TCM
Programme grant and the Swiss National Foundation fellowship.

\end{document}